\documentclass[english,aps, prb, twocolumn]{revtex4-1}
\usepackage[T1]{fontenc}
\usepackage[latin9]{inputenc}
\usepackage{amsmath}
\usepackage{amssymb}
\usepackage{graphicx}
\usepackage{subscript}

\makeatletter

\providecommand{\tabularnewline}{\\}

\@ifundefined{textcolor}{}
{%
 \definecolor{BLACK}{gray}{0}
 \definecolor{WHITE}{gray}{1}
 \definecolor{RED}{rgb}{1,0,0}
 \definecolor{GREEN}{rgb}{0,1,0}
 \definecolor{BLUE}{rgb}{0,0,1}
 \definecolor{CYAN}{cmyk}{1,0,0,0}
 \definecolor{MAGENTA}{cmyk}{0,1,0,0}
 \definecolor{YELLOW}{cmyk}{0,0,1,0}
}

\makeatother

\usepackage{babel}
\begin{document}

\preprint{This line only printed with preprint option}

\title{Evaluation of the importance of spin-orbit couplings in the nonadiabatic
quantum dynamics with quantum fidelity and with its efficient ``on-the-fly''
\textit{ab initio} semiclassical approximation}

\author{Tomá\v{s} Zimmermann}

\author{Ji\v{r}\'{\i} Van\'{\i}\v{c}ek}

\email{jiri.vanicek@epfl.ch}

\selectlanguage{english}%

\affiliation{Laboratory of Theoretical Physical Chemistry, Institut des Sciences
et Ingénierie Chimiques, Ecole Polytechnique Fédérale de Lausanne
(EPFL), CH-1015, Lausanne, Switzerland}
\begin{abstract}
We propose to measure the importance of spin-orbit couplings (SOCs)
in the nonadiabatic molecular quantum dynamics rigorously with quantum
fidelity. To make the criterion practical, quantum fidelity is estimated
efficiently with the multiple-surface dephasing representation (MSDR).
The MSDR is a semiclassical method that includes nuclear quantum effects
through interference of mixed quantum-classical trajectories without
the need for the Hessian of potential energy surfaces. Two variants
of the MSDR are studied, in which the nuclei are propagated either
with the fewest-switches surface hopping or with the locally mean
field dynamics. The fidelity criterion and MSDR are first tested on
one-dimensional model systems amenable to numerically exact quantum
dynamics. Then, the MSDR is combined with ``on-the-fly'' computed
electronic structure to measure the importance of SOCs and nonadiabatic
couplings (NACs) in the photoisomerization dynamics of CH\textsubscript{2}NH\textsubscript{2}\textsuperscript{+}
considering 20 electronic states and in the collision of F + H\textsubscript{2}
considering six electronic states. 
\end{abstract}


\maketitle

\section{Introduction\label{sec:Introduction}}

Nonadiabatic couplings (NACs) originating in the Born-Oppenheimer
separation of motion of electrons and nuclei often play an important
role in the molecular dynamics.\cite{FDisscuss2004} In photochemistry,
NACs are responsible for radiationless decay of the electronically
excited states by a process called internal conversion.\cite{Worth2004,Martinez2010}
In chemical reaction dynamics, they may affect branching ratios of
product channels or allow for reactivity of otherwise nonreactive
states, when the transition from a nonreactive potential energy surface
(PES) to a reactive PES is induced by the motion of nuclei.\cite{Butler1998,Wu2010}
In some cases, transitions between PESs may be induced by another
type of couplings called spin-orbit couplings (SOCs), which couple
the electronic spin with angular momentum.\cite{Marian2012} Relativistic
in origin, SOCs become increasingly important in molecules containing
heavy elements. Nevertheless, SOCs often play a role also in molecules
composed of light elements, e.g., various chromophores or DNA bases.\cite{Gonzalez-Luque2010}
In photochemistry, SOCs cause radiationless transitions between PESs
of different multiplicity called intersystem crossings, which are
responsible for the phosphorescence or emergence of triplet intermediates
in photochemical reactions.\cite{Neiss2003,Griesbeck2004,Abe2004}
Similarly to NACs, SOCs may affect chemical reactions even when no
photoexcitation is involved.\cite{Maiti2004,Lique2011} 

Even though intersystem crossing is typically slower than internal
conversion in molecules composed of light elements, in some cases
both effects may occur on a comparable time scale.\cite{Takezaki1998,Satzger2004,Reichardt2009,Minns2010}
Therefore, there is a need for a rigorous criterion of the importance
of SOCs in accurate quantum nonadiabatic simulations. Knowing in advance
which PESs and which couplings are important for given initial conditions
may speed up a simulation and avoid costly calculations of additional
PESs. Below, we propose and test a criterion of importance of SOCs
which is based on the quantum fidelity.\cite{Peres1984} In a similar
manner, fidelity was already used to measure nondiabaticity\cite{Zimmermann2010}
or nonadiabaticity\cite{MacKenzie2011,Zimmermann2012} of quantum
dynamics. However, direct evaluation of fidelity requires knowledge
of the quantum state evolved with the complete Hamiltonian including
SOCs and consequently is impractical as a guide for quantum simulations.
Fortunately, there exists a fast semiclassical method to compute quantum
fidelity, called the \textit{multiple-surface dephasing representation}
(MSDR).\cite{Zimmermann2010,Zimmermann2012} Evaluation of the importance
of SOCs with the MSDR is the main focus of this work.

Originally, the MSDR was introduced to measure either the nonadiabaticity
or nondiabaticity of the quantum molecular dynamics.\cite{Zimmermann2010,Zimmermann2012}
Roughly speaking, \textquotedblleft{}nondiabaticity\textquotedblright{}
is the difference between the diabatic quantum dynamics (i.e., quantum
dynamics in which the diabatic coordinate couplings between diabatic
electronic surfaces are neglected) and the fully coupled quantum dynamics.
Similarly, \textquotedblleft{}nonadiabaticity\textquotedblright{}
is the difference between the adiabatic quantum dynamics (i.e., quantum
dynamics in which the nonadiabatic momentum couplings between adiabatic
electronic surfaces are neglected) and the fully coupled quantum dynamics.
In both cases the \textquotedblleft{}difference\textquotedblright{}
between the two types of dynamics is measured by the decay of the
overlap of the time-dependent molecular wave functions evolved using
the two types of dynamics. In the same manner, the importance of SOCs
may be evaluated either in: 1) the basis which is diabatic with respect
to SOCs, meaning that the Hamiltonian matrix expressed in this basis
does not contain any spin-orbit-related couplings that depend on the
nuclear momentum, or 2) in the basis which is adiabatic with respect
to SOCs, meaning that the only spin-orbit-related couplings present
in the Hamiltonian depend on the nuclear momentum. The basis which
is diabatic with respect to SOCs is usually more suitable when SOCs
represent only a small perturbation, whereas the basis which is adiabatic
with respect to SOCs is more appropriate in case of strong coupling.
Independently of the adiabaticity or diabaticity with respect to SOCs,
the basis may be adiabatic or diabatic with respect to couplings originating
in the Born-Oppenheimer separation of motion of electrons and nuclei.
Below, we use exclusively the basis which is diabatic with respect
to SOCs. When analyzing the importance of SOCs, we always evaluate
the importance of the full spin-orbit Hamiltonian which may include
diagonal terms as well. Nevertheless, the fidelity criterion and MSDR
may be used also in the fully adiabatic basis of spin-orbit states,
which is employed, for example, in the mixed quantum-classical code
SHARC.\cite{Richter2011} It is important to note that spin-orbit-related
couplings have different meaning in the two basis sets. In the diabatic
basis with respect to SOCs, when the full spin-orbit Hamiltonian is
considered as a perturbation, SOCs reflect faithfully the importance
of the spin-orbit interaction in the system. In the adiabatic basis
with respect to SOCs, on the other hand, the spin-orbit interaction
modifies the PESs themselves and the lack of importance of the offdiagonal
spin-orbit-related couplings does not necessarily mean that the spin-orbit
interaction is unimportant.

The MSDR is a generalization to nonadiabatic dynamics of the dephasing
representation (DR),\cite{Vanicek2003,Vanicek2004,Vanicek2006} derived
for dynamics on a single surface using the Van Vleck propagator or
the linearization of the path integral.\cite{Shi2004} In the single-surface
setting, the DR is closely related to the semiclassical perturbation
approximation of Smith, Hubbard, and Miller\cite{Smith1978,Hubbard1983}
and to the phase averaging of Mukamel.\cite{Mukamel1982} Its applications
include evaluations of the stability of quantum dynamics\cite{Vanicek2004,Vanicek2006,Gorin2006,Wisniacki2010,Garcia-mata2011b}
and calculations of electronic spectra.\cite{Mukamel1982,Rost1995,Li1996,Egorov1998,Shi2005,Wehrle2011,Sulc2012}

While the MSDR is not a method for general dynamics, it can be used
to compute all quantities which can be expressed in terms of quantum
fidelity amplitude. Weaker generality and semiclassical nature give
the MSDR some advantages in comparison to general methods for nonadiabatic
dynamics. 

The main advantage of the MSDR compared to wave packet methods is
that the number of trajectories needed for convergence of MSDR does
not scale exponentially with the number of degrees of freedom.\cite{Mollica2011a}
Therefore, the MSDR may be applied to problems with dimensionality
beyond the scope of even the most advanced methods for nonadiabatic
quantum dynamics such as the multi-configuration time-dependent Hartree
methods.\cite{Meyer2009,Worth2008,Burghardt2008,Blancafort2011} Similarly
to other semiclassical methods,\cite{Martinez1996,Yang2009,Burant2002,Kondorskiy2004,Wu2005,Sun1997,Miller2009,Heller2002}
the MSDR includes some quantum effects on the nuclear motion. The
advantage of MSDR in comparison to most other semiclassical approaches
is that the MSDR does not require the Hessian of the potential energy,
which is often the most expensive part of semiclassical calculations
(see, e.g., Ref. \onlinecite{Ceotto2009a}). In terms of the computational
cost per trajectory, the MSDR thus falls into the category of methods
in which nuclei are treated classically such as surface hopping methods,\cite{Tully1990,Nielsen2000,Shenvi2009}
and other methods based on the mixed quantum-classical Liouville equation,\cite{Donoso1998,Kapral1999,Horenko2002,Ando2003,Horenko2004,Micha2004,MacKernan2008,Bonella2010,Bousquet2011}
or methods obtained by linearization of the path integral expression
for the quantum propagator.\cite{Dunkel2008,Huo2011} (Costs of current
implementations of the MSDR are essentially the costs of a mean field
or surface hopping dynamics. Nevertheless, in principle, the MSDR
may be combined with a propagation scheme based directly on the mixed
quantum-classical Liouville equation, which is typically more expensive
but also more accurate.)

The outline of the paper is as follows: Section\,\ref{sec:Theory}
starts by introducing the fidelity criterion of importance of SOCs
and by defining the spin-orbit Hamiltonian. This is followed by a
brief derivation of the MSDR, and the description of the propagation
scheme and the algorithm. The section ends with the computational
details. In Section\,\ref{sec:Results}, the fidelity criterion and
MSDR are tested using one-dimensional model systems which allow for
numerically exact quantum solution. Subsequently, the MSDR is combined
with ``on-the-fly'' computed \textit{ab initio} electronic structure
and applied to evaluate the importance of SOCs and NACs in the photoisomerization
of CH\textsubscript{2}NH\textsubscript{2}\textsuperscript{+} and
in the collision of F + H\textsubscript{2}. Section\,\ref{sec:Discussion-and-Conclusions}
concludes the paper.

\section{Theory\label{sec:Theory}}

\subsection{Fidelity as a measure of importance of spin-orbit coupling terms}

Following our work on the nondiabaticity\cite{Zimmermann2010} and
nonadiabaticity\cite{Zimmermann2012} of the molecular dynamics we
base the quantitative criterion of importance of SOCs on quantum fidelity
$F_{\mathrm{QM}}$ between molecular quantum states propagated with
and without SOCs. More precisely, 
\begin{equation}
F_{\mathrm{QM}}(t)=|f_{\mathrm{QM}}(t)|^{2}=\vert\langle\psi^{0}(t)\vert\psi^{\epsilon}(t)\rangle\vert^{2},\label{eq:Fidelity_definition}
\end{equation}
where $f_{\mathrm{QM}}$ is the quantum fidelity amplitude, $\left|\psi^{0}(t)\right\rangle =e^{-i\mathbf{\hat{H}}^{0}t/\hbar}\left|\psi(0)\right\rangle $
is the quantum state of the molecule evolved using the nonadiabatic
Hamiltonian $\mathbf{\hat{H}}^{0}$ which does not include SOCs of
interest and $\vert\psi^{\epsilon}(t)\rangle=e^{-i\mathbf{\hat{H}}^{\epsilon}t/\hbar}\left|\psi(0)\right\rangle $
is the quantum state evolved using the fully coupled nonadiabatic
Hamiltonian $\mathbf{\hat{H}}^{\epsilon}=\mathbf{\hat{H}}^{0}+\epsilon\hat{\mathbf{V}}$.
In the last expression, $\hat{\mathbf{V}}=\mathbf{\hat{H}}^{\text{SO}}$
contains SOCs of interest and $\epsilon$ controls the extent of perturbation.
(\textbf{Bold} face denotes $n\times n$ matrices acting on the Hilbert
space spanned by $n$ electronic states, hat $\hat{}$ denotes nuclear
operators. In general, the superscript $0$ or $\epsilon$ designates
the Hamiltonian with which the object was propagated.) When $F_{\mathrm{QM}}\approx1$,
$\left|\psi^{0}(t)\right\rangle $ is close to $\left|\psi^{\epsilon}(t)\right\rangle $
and SOCs do not influence the dynamics significantly. On the other
hand, when $F_{\mathrm{QM}}\ll1$, SOCs are important and should be
taken into account in an accurate quantum calculation. The advantage
of the fidelity criterion is that, in addition to population transfer
between PESs (which is a standard dynamical measure of the importance
of couplings), fidelity can detect subtle effects on the dynamics
caused by the displacement and interference on a single PES (see Ref.\,\onlinecite{Zimmermann2012}
for details).

\subsection{Spin-orbit Hamiltonian}

In principle, any spin-orbit coupling Hamiltonian $\mathbf{\hat{H}}^{\text{SO}}$
may be used. In this work, model potentials are used in one-dimensional
systems. In case of the photoisomerization of CH\textsubscript{2}NH\textsubscript{2}\textsuperscript{+}
and collision of F + H\textsubscript{2}, elements of $\mathbf{\hat{H}}^{\text{SO}}$
are computed with the Breit-Pauli Hamiltonian\cite{Berning2000}
\begin{eqnarray}
\hat{H}^{\mathrm{BP}} & = & \sum_{j,J}\frac{Z_{J}\left(\hat{q}_{jJ}\times\hat{p}_{j}\right)\cdot\hat{s}_{j}}{2c^{2}\left|q_{jJ}\right|^{3}}\nonumber \\
 &  & -\sum_{j\neq k}\left[\frac{\left(\hat{q}_{kj}\times\hat{p}_{j}\right)\cdot\hat{s}_{j}}{2c^{2}\left|q_{jk}\right|^{3}}+\frac{\left(\hat{q}_{jk}\times\hat{p}_{k}\right)\cdot\hat{s}_{j}}{c^{2}\left|q_{jk}\right|^{3}}\right],\label{eq:BP_Hamiltonian}
\end{eqnarray}
where $\hat{q}_{jJ}=(\hat{q}_{j}-\hat{Q}_{J})$ is the difference
of position vectors of electron $j$ and nucleus $J$, $\hat{q}_{jk}=(\hat{q}_{j}-\hat{q}_{k})$
is the difference of position vectors of of electron $j$ and $k$,
$\hat{p}_{j}$ is the momentum of electron $j$, and $\hat{s}_{j}$
is the spin of electron $j$. Exceptionally, since $\hat{H}^{\mathrm{BP}}$
is not expressed in any basis yet, in Eq.\,\eqref{eq:BP_Hamiltonian}
(and only there) hats denote both electronic and nuclear operators.

\subsection{MSDR}

Here we will only summarize the theory of MSDR; the full derivation
may be found in Ref.\,\onlinecite{Zimmermann2012}. The derivation
starts with a quantum fidelity amplitude expression generalized to
the Hilbert space given by the tensor product $\mathbb{C}^{n}\otimes\mathcal{L}^{2}\left(\mathbb{R}^{D}\right)$:\cite{Vanicek2006}
\begin{equation}
f_{\mathrm{QM}}\left(t\right)=\mathrm{Tr}\left(e^{-i\mathbf{\hat{H}}^{\epsilon}t/\hbar}\cdot\mathbf{\mathbf{\boldsymbol{\hat{\rho}}}}^{\text{init}}\cdot e^{+i\mathbf{\hat{H}}^{0}t/\hbar}\right),\label{eq:f_rho}
\end{equation}
 where $\boldsymbol{\hat{\rho}}^{\text{init}}$ is the density operator
of the initial state. Expressing $f_{\mathrm{QM}}$ in the interaction
picture and partially Wigner transforming\cite{Wigner1932} the resulting
equation over nuclear degrees of freedom yields an alternative exact
expression 
\begin{align}
f_{\mathrm{QM}}\left(t\right) & =h^{-D}\mathrm{Tr}_{e}\int d^{2D}X\boldsymbol{\rho}_{\text{W}}^{\text{init}}\left(X\right)\cdot\left({\cal T}e^{-i\epsilon\int_{0}^{t}\mathbf{\hat{V}}^{\text{I}}\left(t'\right)dt'/\hbar}\right)_{\text{W}},\label{eq:f_int_pict_Wigner}
\end{align}
where $X$ denotes the point $(Q,P)$ in the $2D$-dimensional nuclear
phase space, $\mathrm{Tr}_{e}$ is the trace over electronic degrees
of freedom, and $\mathbf{\hat{V}}^{\text{I}}\left(t\right)$ is the
perturbation in the interaction picture given by $\mathbf{\hat{H}}^{0}$.
The first approximation consists in replacing the Wigner transform
of a product of operators in the Taylor expansion of the time-ordered
exponential in Eq.\,\eqref{eq:f_int_pict_Wigner} by the product
of Wigner transforms of these operators. Recognizing the resulting
sum as an Taylor expansion of another exponential, this approximation
can be expressed succinctly as 
\begin{equation}
\left({\cal T}e^{-i\epsilon\int_{0}^{t}\mathbf{\hat{V}}^{\text{I}}\left(t'\right)dt'/\hbar}\right)_{\text{W}}\simeq{\cal T}e^{-i\epsilon\int_{0}^{t}\mathbf{V}_{\text{W}}^{\text{I}}\left(X,t'\right)dt'/\hbar}.\label{eq:approx_product_Wigner_transf}
\end{equation}
To evaluate this expression, the time evolution of $\mathbf{V}_{\text{W}}^{\text{I}}\left(X,t\right)$
has to be known. The second approximation involves replacing the exact
evolution by a mixed quantum-classical (MQC) propagation scheme described
below, leading to the final expression for the MSDR of fidelity amplitude,
\begin{multline}
f_{\text{MSDR}}(t)=\left\langle {\cal T}e^{-i\epsilon\int_{0}^{t}\mathbf{V}_{\text{W,MQC}}^{\text{I}}\left(X,t'\right)dt'/\hbar}\right\rangle _{\boldsymbol{\rho}_{\text{W}}^{\text{init}}\left(X\right)}\\
=h^{-D}\mathrm{Tr}_{e}\int d^{2D}X\boldsymbol{\rho}_{\text{W}}^{\text{init}}(X)\cdot{\cal T}e^{-i\epsilon\int_{0}^{t}\mathbf{V}_{\text{W,MQC}}^{\text{I}}\left(X,t'\right)dt'/\hbar}.\label{eq:f_DR_D}
\end{multline}

\subsection{Propagation scheme}

The MQC equation\cite{Aleksandrov1981,Boucher1988,Martens1997,Prezhdo1997,Kapral1999,Caro1999,Shi2004}
for the evolution of the density matrix is given by
\begin{multline}
\frac{\partial\boldsymbol{\rho}_{\text{W,MQC}}}{\partial t}=-\frac{i}{\hbar}\left[\mathbf{H}_{\text{W}},\boldsymbol{\rho}_{\text{W,MQC}}\right]\\
+\frac{1}{2}\left(\left\{ \mathbf{H}_{\text{W}},\boldsymbol{\rho}_{\text{W,MQC}}\right\} -\left\{ \boldsymbol{\rho}_{\text{W,MQC}},\mathbf{H}_{\mathrm{\text{W}}}\right\} \right),\label{eq:mixed_Liouville}
\end{multline}
where the explicit dependence of $\boldsymbol{\rho}_{\text{W,MQC}}$
on time and on the nuclear phase-space coordinate $X$ was omitted
for clarity. The propagation equation for $\mathbf{V}_{\text{W,MQC}}^{\text{I}}\left(X,t\right)$,
which is the last ingredient needed in our method, differs from Eq.\,\eqref{eq:mixed_Liouville}
only by the sign of the time derivative. Therefore Eq.\,\eqref{eq:f_DR_D}
together with Eq.\,\eqref{eq:mixed_Liouville} define the MSDR. Several
numerical approaches exist that solve Eq.\,\eqref{eq:mixed_Liouville}
in terms of ``classical'' trajectories $X(t)$. However, since trajectory-based
methods for solving Eq.\,\eqref{eq:mixed_Liouville} are still relatively
complicated, the MSDR is in practice implemented using one of two
schemes which further approximate Eq.\,\eqref{eq:mixed_Liouville}.
The common feature of the two approximations is that all elements
of $\boldsymbol{\rho}_{\text{W}}\left(X,t\right)$ are propagated
using the same PES (which may, nevertheless, differ for different
trajectories). The first approach\cite{Zimmermann2012} approximates
Eq.\,\eqref{eq:mixed_Liouville} as 
\begin{multline}
\frac{\partial\boldsymbol{\rho}_{\text{W,LMFD}}}{\partial t}=-\frac{i}{\hbar}\left[\mathbf{H}_{\text{W}},\boldsymbol{\rho}_{\text{W,LMFD}}\right]\\
+\frac{\partial\boldsymbol{\rho}_{\text{W,LMFD}}}{\partial P}\left\langle \frac{\partial\mathbf{H}_{\text{W}}}{\partial Q}\right\rangle _{e}-\frac{\partial\boldsymbol{\rho}_{\text{W,LMFD}}}{\partial Q}\frac{P}{M},\label{eq:Liouville_diab_local-1}
\end{multline}
where $\left\langle \mathbf{A}\right\rangle _{e}=\mathrm{Tr}_{e}\left(\boldsymbol{\rho}_{e}\cdot\mathbf{A}\right)$
is a partial average of $\mathbf{A}$ over the electronic subspace.
This approach was called the \textit{locally mean field dynamics}
(LMFD) in Ref.\,\onlinecite{Zimmermann2012}, where it was derived
simply by invoking the locally mean field approximation. As pointed
out in Ref.\,\onlinecite{Zimmermann2012}, the LMFD turns out to
be nothing else than the dynamics of a swarm of trajectories, each
of which is propagated with the Ehrenfest dynamics. The second approach
employs the physically motivated fewest-switches surface hopping (FSSH)
scheme,\cite{Tully1990} where the matrix elements of the density
operator $\boldsymbol{\rho}_{\text{W,FSSH}}$ are (in the adiabatic
basis) computed using Eq. (11) of Ref.\,\onlinecite{Tully1990}.

\subsection{Algorithm}

The details of the general implementation of the MSDR algorithm are
given in Ref.\,\onlinecite{Zimmermann2012}. In this work, we consider
only initial states for which the ``conditional'' electronic density
matrix is pure for all $X$ and hence the initial density matrix can
be written as the tensor product
\begin{equation}
\boldsymbol{\rho}_{\text{W}}^{\text{init}}\left(X\right)=\rho^{\text{init}}\left(X\right)\mathbf{c^{\mathrm{init}}}\left(X\right)\otimes\mathbf{c}^{\mathrm{init}}\left(X\right)^{\dagger}\label{eq:init_density_pure}
\end{equation}
where $\rho^{\text{init}}\left(X\right):=\mathrm{Tr}_{e}\boldsymbol{\rho}_{\mathrm{W}}^{\text{init}}\left(X\right)$
and $\mathbf{c^{\mathrm{init}}}\left(X\right)$ is the initial electronic
wave function for nuclei located at $X$. In practice, $N_{\text{traj}}$
phase space points are sampled from a scalar nuclear density $\rho^{\text{init}}(X)$
and a vector $\mathbf{c^{\mathrm{init}}}\left(X\right)$ is attributed
to each of the generated phase space points. In the case of LMFD,
this determines the initial condition completely. In the case of FSSH,
one also needs to select the initial surface randomly and separately
for each trajectory according to the following prescription: For a
trajectory starting at $X$, the probability for its initial surface
to be surface $j$ is given by $\left|c_{j}^{\text{init}}(X)\right|^{2}$.
The trajectories are then propagated with $\mathbf{H}_{\text{W}}^{0}$
{[}because, ultimately, we propagate $\mathbf{V}_{\text{W,MQC}}^{\text{I}}\left(X,t\right)$
in the interaction picture given by $\mathbf{H}_{\text{W}}^{0}${]}
by using either the LMFD or FSSH dynamics. It is advantageous that
for states \eqref{eq:init_density_pure}, when the LMFD or FSSH dynamics
is used, Eq.\,\eqref{eq:f_DR_D} simplifies to the weighted phase
space average 
\begin{equation}
f_{\text{MSDR}}\left(t\right)=\left\langle \mathbf{c}^{0}\left(X,t\right)^{\dagger}\cdot\mathbf{c}^{\epsilon}\left(X,t\right)\right\rangle _{\rho^{\text{init}}\left(X\right)}.\label{eq:f_DR_D_pure_states}
\end{equation}
In Eq.\,\eqref{eq:f_DR_D_pure_states}, the wave function $\mathbf{c}^{0}\left(X,t\right)$
is obtained automatically from the LMFD or FSSH dynamics. Analogously,
$\mathbf{c}^{\epsilon}\left(X,t\right)$ solves the Schr\"odinger
equation 
\begin{equation}
\frac{\partial\mathbf{c}^{\epsilon}\left(X,t\right)}{\partial t}=-\frac{i}{\hbar}\mathbf{H}_{\text{W}}^{\epsilon}(X^{0}(t))\cdot\mathbf{c}^{\epsilon}\left(X,t\right)\label{eq:ceps_schroedinger}
\end{equation}
 for a single discrete electronic degree of freedom with a time-dependent
Hamiltonian $\mathbf{H}_{\text{W}}^{\epsilon}(t):=\mathbf{H}_{\text{W}}^{\epsilon}(X^{0}(t))$
(i.e., in the Lagrangian reference frame given by $\mathbf{H}_{\text{W}}^{0}$)
and with the initial condition $\mathbf{c}^{\epsilon}\left(X,0\right)=\mathbf{c}^{\mathrm{init}}\left(X\right)$.
In Eq.\,\eqref{eq:ceps_schroedinger}, $X^{0}(t)$ is the phase space
point resulting from the evolution of the initial phase space point
$X$ for time $t$ with $\mathbf{H}_{\text{W}}^{0}$, using either
the LMFD or FSSH dynamics.

\subsection{Computational details}

All quantum calculations were performed using the second-order split-operator
algorithm.\cite{Feit1983} The LMFD or FSSH dynamics were done using
the second-order symplectic Verlet integrator.\cite{Verlet1967} The
Schr\"odinger equation for the discrete ``electronic'' degree of
freedom was solved using the unitary propagator $\mathbf{U}(X,t,t+\Delta t)=e^{-i\mathbf{H}_{\text{W}}\left(X\right)\Delta t/\hbar}$.
\textit{Ab initio} calculations were performed using Molpro 2010.1\cite{MOLPRO}
and Columbus 5.9.2.\cite{COLUMBUS5_9_2}

\section{Results\label{sec:Results}}

\subsection{Comparison of the MSDR with the numerically exact quantum dynamics}

\begin{figure*}
\includegraphics[width=\textwidth]{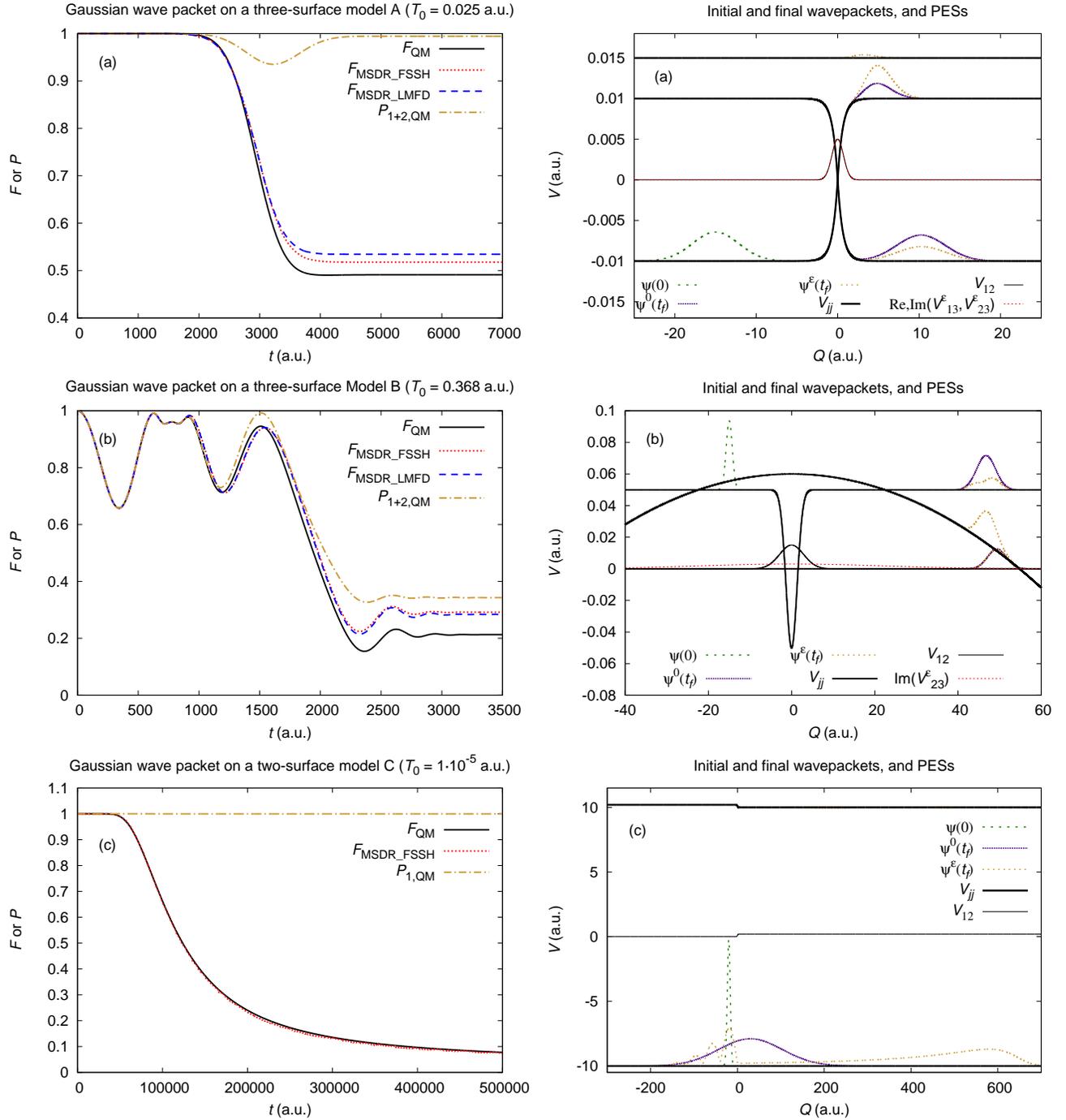}

\caption{Importance of SOCs in the nonadiabatic dynamics: comparison with the
numerically exact quantum dynamics on model systems. Left panels 
compare the quantum fidelity ($F_{\mathrm{QM}}$) with the MSDR ($F_{\mathrm{MSDR}}$)
and with the quantum survival probability ($P_{\mathrm{QM}}$). Right
panels show the corresponding diabatic PESs and couplings $V_{jj}(Q)$
as well as initial {[}$\psi(0)${]} and final {[}$\psi^{0}(t_{f})$
and $\psi^{\epsilon}(t_{f})${]} wave functions evolved with $\mathbf{\hat{H}}^{0}$
and $\mathbf{\hat{H}}^{\epsilon}$, respectively. In all three models,
the initial state was the GWP $\psi(q)=\frac{1}{\sigma\sqrt{\pi}}\exp\left[-(Q-Q_{0})^{2}/2\sigma^{2}+P_{0}(Q-Q_{o})/\hbar\right]$
with the mass equal to 2000 $\mathrm{a.u.}$ (a)\,Coupling terms
$V_{13}=V_{23}$, containing both real and imaginary parts, result
in a very low decay of $P_{\mathrm{QM}}$ which does not correspond
to the substantial decay of $F_{\mathrm{QM}}$. Only the lowest-energy
PES was occupied initially by a GWP with $Q_{0}=-15$ $\mathrm{a.u.}$,
$P_{0}=10$ $\mathrm{a.u.}$, and $\sigma=2.83$ $\mathrm{a.u.}$
(b)\,Extensive but relatively weak purely imaginary spin-orbit coupling
term causes considerable decay of both $F_{\mathrm{QM}}$ and $P_{\mathrm{QM}}$.
Only the second lowest-energy PES was occupied initially by a GWP
with $Q_{0}=-15$ $\mathrm{a.u.}$, $P_{0}=38.36$ $\mathrm{a.u.}$,
and $\sigma=0.74$ $\mathrm{a.u.}$ (c)\,The coupling term $V_{12}$
causes decay of $F_{\mathrm{QM}}$ towards zero without affecting
$P_{\mathrm{QM}}$. Only the lowest-energy PES was occupied initially
by a GWP with $Q_{0}=-20$ $\mathrm{a.u.}$, $P_{0}=0.2$ $\mathrm{a.u.}$,
and $\sigma=3.41$ $\mathrm{a.u.}$ \label{fig:comparison_with_quantum}}
\end{figure*}

In order to test the utility of quantum fidelity as a measure of the
importance of SOCs and the performance of the MSDR in comparison to
the numerically exact quantum dynamics we use three one-dimensional
model systems. Results are shown in Fig.\,\ref{fig:comparison_with_quantum}. 

The PESs and couplings of model A are shown in the right part of Fig.\,\ref{fig:comparison_with_quantum}\,(a).
Both Hamiltonians $\mathbf{\hat{H}}^{0}$ and $\mathbf{\hat{H}}^{\epsilon}$
are expressed in the diabatic basis with respect to both SOCs and
couplings originating in the Born-Oppenheimer separation of motion
of electrons and nuclei. Both Hamiltonians contain three PESs, of
which the lower two are identical to the PESs of Tully's single avoided
crossing model.\cite{Tully1990} The third PES is flat with constant
energy $E=0.15\,\mathrm{a.u.}$ In both Hamiltonians, the lower two
PESs are coupled by the same coupling term $V_{12}$ as in the original
single avoided crossing model. Additionally, in $\mathbf{\hat{H}}^{\epsilon}$
(but not in $\mathbf{\hat{H}}^{0}$), the highest PES is coupled to
the lower two PESs with $V_{13}=V_{31}^{*}=V_{23}=V_{32}^{*}=\left(1+i\right)\left[C\exp\left(-DQ^{2}\right)\right],$
where $C=0.005$ and $D=1.0$. A more general, complex form was chosen
to emulate spin-orbit coupling terms, which may also be complex-valued.
The initial state is a Gaussian wave packet (GWP) located on the lowest
energy PES with the mean kinetic energy $T_{0}=0.025$ $\mathrm{a.u.}$
As can be seen in Fig.\,\ref{fig:comparison_with_quantum}\,(a),
quantum fidelity $F_{\mathrm{QM}}$ decays substantially (by more
than 50\%), signifying the importance of $V_{13}$ and $V_{23}$ in
the dynamics. The decay of $F_{\mathrm{QM}}$ is accurately reproduced
by the MSDR using both the LMFD and FSSH dynamics. On the other hand,
the decay of the survival probability $P_{1+2,\text{QM}}=1-P_{3,\text{QM}}$
is very small and the final probability $P_{3,\text{QM}}$ of finding
the system on the third PES is less than 1\%. This demonstrates that
the survival probability $P_{1+2,\text{QM}}$ may be a poor measure
of the importance of couplings between PESs. More refined criteria,
such as the fidelity criterion, are needed to measure the influence
of SOCs. (Note that in this specific case, strong effects of couplings
of the first two PESs to the third PES may be inferred from considering
the probabilities P$_{1}$ and P$_{2}$ separately. However, for that
it is necessary to run the dynamics twice - once with couplings to
the third surface and once without the couplings. To use the fidelity
criterion approximated by the MSDR only one simulation is sufficient.)

Model B shown in Fig.\,\ref{fig:comparison_with_quantum}\,(b) is
based on the two-surface Tully's double avoided crossing model.\cite{Tully1990}
It is also expressed in the diabatic basis with respect to both SOCs
and couplings originating in the Born-Oppenheimer separation of motion
of electrons and nuclei. The third additional surface is described
by the equation $V_{33}=FQ^{2}+G,$ where $F=-2\cdot10^{-5}$ and
$G=0.06$. The coupling term $V_{23}=V_{32}^{*}=iI\exp\left(-JQ^{2}\right)$
(with $I=0.003$ and $J=0.001$) is present only in $\mathbf{\hat{H}}^{\epsilon}$.
As can be seen in the right part of Fig.\,\ref{fig:comparison_with_quantum}\,(b),
the coupling term $V_{23}$ is relatively weak but widely spread,
resembling closely SOCs typically seen in molecular dynamics. The
initial state is a GWP located on the second lowest-energy PES with
mean kinetic energy $T_{0}=0.368$ $\mathrm{a.u.}$ Again, the decay
of $F_{\mathrm{QM}}$ is closely followed by both implementations
of the MSDR. This time, even $P_{1+2,\text{QM}}$ follows $F_{\mathrm{QM}}$
relatively well. Still, the approximate MSDR method gives a slightly
more accurate picture of the influence of SOCs on the dynamics.

Model C {[}Fig.\,\ref{fig:comparison_with_quantum}\,(c){]}, inspired
by Tully's extended coupling model,\cite{Tully1990} demonstrates
that the fidelity criterion detects the importance of SOCs even in
cases where no significant transition of the probability density between
surfaces occurs. The PESs and couplings are given by
\begin{eqnarray*}
V_{11}\left(Q\right) & = & -A\\
V_{22}\left(Q\geq0\right) & = & A+B\exp\left(-CQ\right)\\
V_{22}\left(Q<0\right) & = & A+B\left[2-\exp\left(CQ\right)\right]\\
V_{12}\left(Q\geq0\right) & = & B\left[2-\exp\left(-CQ\right)\right]\\
V_{12}\left(Q<0\right) & = & B\exp\left(CQ\right),
\end{eqnarray*}
where $A=10$, $B=0.1,$ and $C=0.9$. The real-valued coupling term
$V_{12}$ is present in $\mathbf{\hat{H}}^{\epsilon}$ but not in
$\mathbf{\hat{H}}^{0}$. (Note that when the coupling term is purely
imaginary, the results are exactly identical.) As can be seen in Fig.\,\ref{fig:comparison_with_quantum}\,(c),
for a low energy wavepacket the wavefunctions propagated with $\mathbf{\hat{H}}^{0}$
and $\mathbf{\hat{H}}^{\epsilon}$ are very different. Indeed, fidelity
decays towards zero. On the other hand, the survival probability $P_{1}$
never decays by more than $6\cdot10^{-5}$. Even though the model
is slightly artificial (in that the PESs are considerably coupled,
despite a large energy difference between them), it clearly demonstrates
the plausibility of situations in which the dynamical importance of
SOCs would be undetectable if measured with the survival probability
as a criterion. Finally, Fig.\,\ref{fig:comparison_with_quantum}\,(c)
shows that the MSDR still closely follows the quantum result. {[}Since
$\mathbf{\hat{H}}^{0}$ is uncoupled, both the LMFD and FSSH dynamics
are equivalent. Therefore, only the FSSH result is shown in Fig.\,\ref{fig:comparison_with_quantum}\,(c).{]}

\subsection{On-the-fly \textit{ab initio} photoisomerization dynamics of CH\textsubscript{2}NH\textsubscript{2}\textsuperscript{+}}

As the first on-the-fly \textit{ab initio} application of the MSDR,
we examined the importance of SOCs and NACs in the quantum dynamics
of the second excited singlet state of CH\textsubscript{2}NH\textsubscript{2}\textsuperscript{+}
using the SA5,5-CASSCF(6,4)/6-31G{*} method. The time step of the
FSSH dynamics was equal to $\sim0.2$ $\mathrm{fs}$. In total, five
lowest-energy singlet states and 15 lowest-energy triplet states (five
states for each $M_{S}=1,0,-1$) were considered in the simulation.
To evaluate the nonadiabaticity, the unperturbed Hamiltonian $\mathbf{\hat{H}}^{0}$
did not contain any couplings between surfaces and the perturbed Hamitonian
$\mathbf{\hat{H}}^{\epsilon}$ contained NACs between singlet states.
To evaluate the importance of SOCs, $\mathbf{\hat{H}}^{0}$ contained
only NACs between singlet states while $\mathbf{\hat{H}}^{\epsilon}$
contained NACs between singlet states and SOCs between all 20 considered
surfaces. The initial state was the vibrational ground state of the
ground electronic PES computed in the harmonic approximation. (The
harmonic approximation was used only to compute the initial state,
and not for the propagation itself.) This wave packet was placed on
the second excited PES (in the basis diabatic with respect to SOCs
and adiabatic with respect to couplings originating in the Born-Oppenheimer
separation of motion of electrons and nuclei). Immediately after excitation,
the system quickly decayed to the first excited state and subsequently
to the ground state. Three main pathways were observed during the
first 100 $\mathrm{fs}$, in agreement with experimental observations
and previous calculations:\cite{Donchi1988,Barbatti2006,Barbatti2007}
1) photoisomerization leading to the hot ground state CH\textsubscript{2}NH\textsubscript{2}\textsuperscript{+}
2) bi-pyramidalisation leading to dissociation into CH\textsubscript{2}\textsuperscript{+}
and NH\textsubscript{2} and 3) release of H\textsubscript{2}. 

\begin{figure}
\begin{tabular}{c}
\includegraphics[scale=0.68]{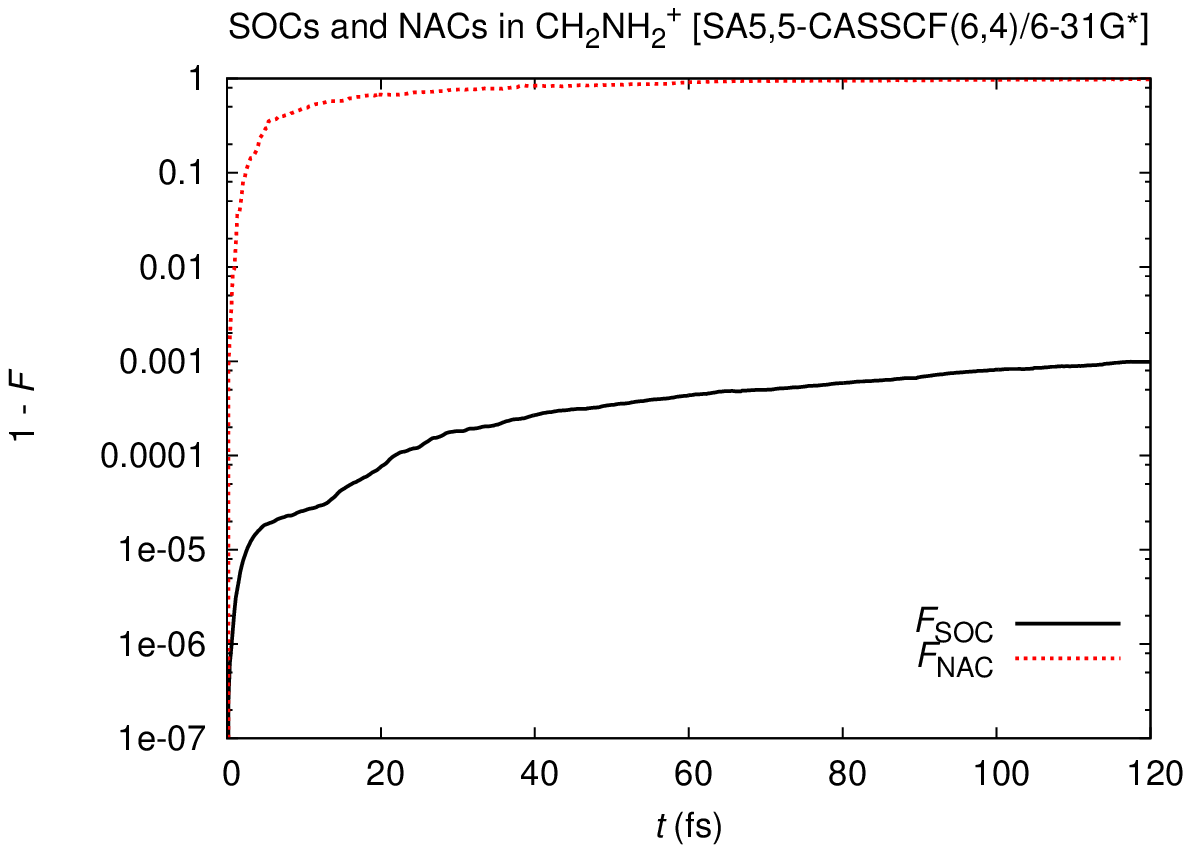}\tabularnewline
\includegraphics[scale=0.2]{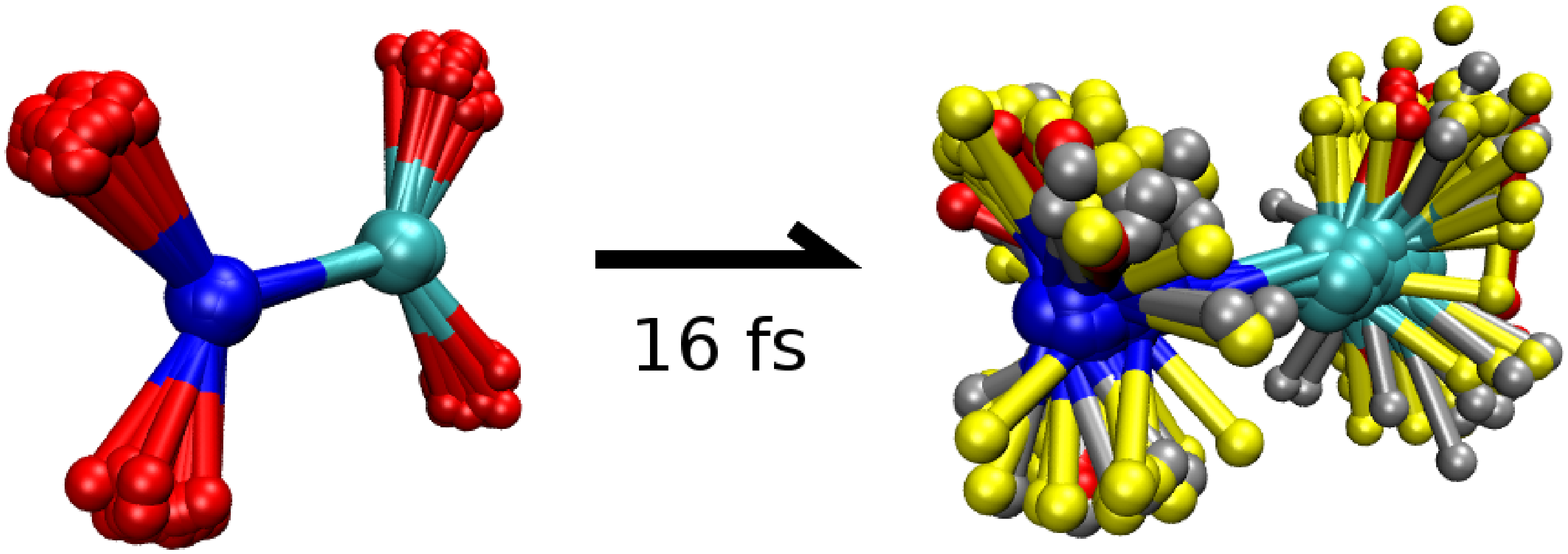}\tabularnewline
\end{tabular}

\caption{Importance of the SOCs and NACs in the dynamics starting on the second
excited electronic state of CH\textsubscript{2}NH\textsubscript{2}\textsuperscript{+}
evaluated with the MSDR combined with the FSSH dynamics. Top: Fast
decay of $F_{\text{NAC}}$ signifies the importance of NACs, whereas
a very slow decay of $F_{\text{SOC}}$ shows that SOCs may be safely
neglected. Bottom: Swarm of 72 trajectories representing the wave
packet during the photoisomerization dynamics of CH\textsubscript{2}NH\textsubscript{2}\textsuperscript{+}
immediately after excitation and $\sim16$\,$\mathrm{fs}$ after
excitation. Configurations on the first, second, and third lowest-energy
excited states are depicted with yellow, red, and pink hydrogen atoms,
respectively. Configurations on the ground state have grey hydrogens.
\label{fig:formaldiminium_SO_nonad}}
\end{figure}

Fast decay of fidelity $F_{\text{NAC}}$, which can be seen in Fig.\,\ref{fig:formaldiminium_SO_nonad},
demonstrates the vast importance of NACs in the quantum dynamics of
photoexcited CH\textsubscript{2}NH\textsubscript{2}\textsuperscript{+}.
On the other hand, SOCs are not important in this dynamics as signified
by the very slow decay of fidelity $F_{\text{SOC}}$. 

\subsection{On-the-fly \textit{ab initio} collision of F + H\textsubscript{2}}

The collision of F + H\textsubscript{2} represents a system where
SOCs are known to be important at least at low energies.\cite{Lique2008,Lique2011}
In order to check the ability of the MSDR to detect the importance
of SOCs, we have applied it together with on-the-fly computed electronic
structure using the SA3-CASSCF(9,10)/6-31+G{*} method. The (9,10)
active space was chosen because it is known to produce qualitatively
correct smooth PESs for any orientation of H\textsubscript{2} axis.\cite{BauschlicherJr.1989}
The time step of the FSSH dynamics was equal to $\sim0.2$ $\mathrm{fs}$.
Six lowest-energy doublet states were considered in the simulation
(three states with $M_{S}=1/2$ and three states with $M_{S}=-1/2$).
The initial state was placed either on the electronic ground state
or on the first excited state with $M_{S}=1/2$ in the basis diabatic
with respect to SOCs and adiabatic with respect to couplings originating
in the Born-Oppenheimer separation of motion of electrons and nuclei.
(The degeneracy of p states of F was removed by H\textsubscript{2}
at a distance of $\sim6$ Å.) The molecule of H\textsubscript{2}
was in the vibrational and rotational ground state with uniformly
random orientation of the molecular axis. Translational degrees of
freedom of F and H\textsubscript{2} were sampled from GWPs with the
mean total kinetic energy of the collision $E_{k}=0.7$ $\mathrm{kcal/mol}$.
At such a low energy, both considered collisions were nonreactive. 

\begin{figure}
\begin{tabular}{c}
\includegraphics[scale=0.68]{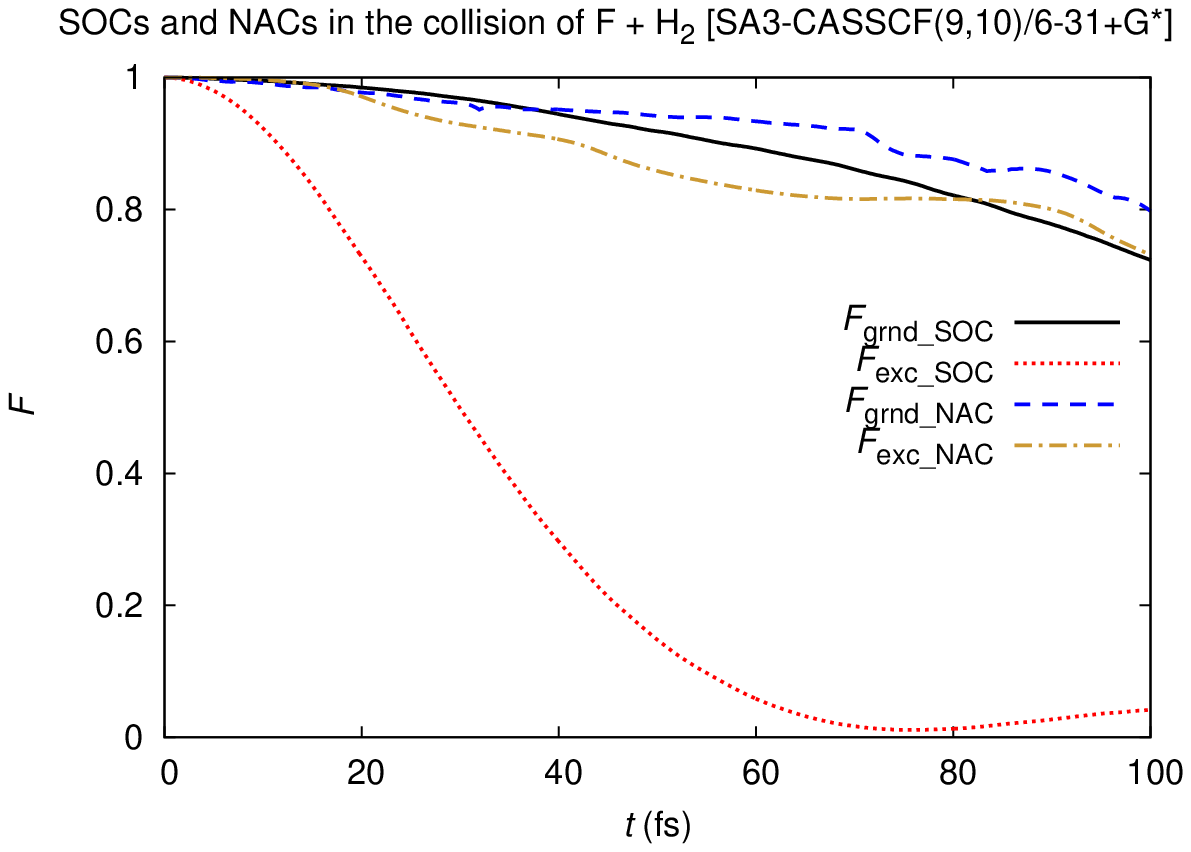}\tabularnewline
\includegraphics[scale=0.19]{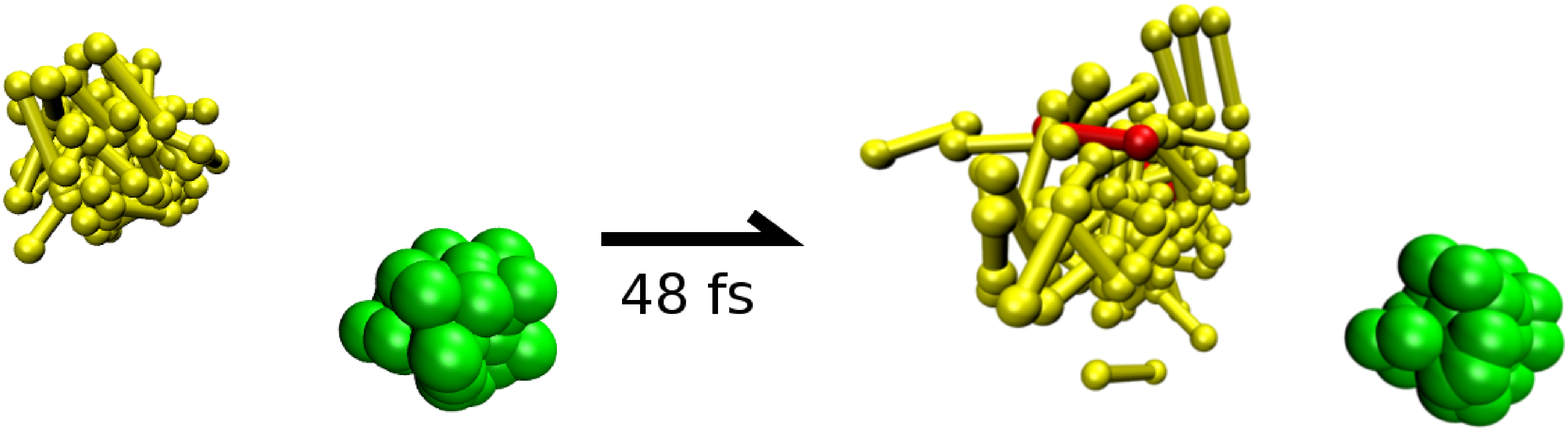}\tabularnewline
\end{tabular}

\caption{Importance of SOCs and NACs in the collision of F + H\textsubscript{2}
evaluated with the MSDR combined with the FSSH dynamics. Top: The
decays of $F_{\text{grnd\_SOC}}$ and $F_{\text{exc\_SOC}}$ signify
the importance of SOCs, especially in the collision initiated on the
first excited electronic state. NACs for both excited and ground state
dynamics are as important as SOCs for the ground state dynamics. Bottom:
Swarm of 64 trajectories representing the wave packet (initially located
on the first excited PES) during the collision of F + H\textsubscript{2}
at $t=0$ $\mathrm{fs}$ and $t\sim48$ $\mathrm{fs}$. Configurations
on the first and second lowest-energy excited doublet states are depicted
with yellow and red hydrogen atoms, respectively. Configurations on
the ground state have grey hydrogens.\label{fig:F_H2_SO_nonad}}
\end{figure}

As can be seen in Fig.\,\ref{fig:F_H2_SO_nonad}, fidelity due to
SOCs decays significantly in both cases. Relatively fast decay of
fidelity in the case of electronically excited initial state signifies
importance of SOCs on the collision dynamics. In the case of the ground
state collision, fidelity decays more slowly and SOCs may probably
be neglected in less accurate simulations. Nevertheless, the decay
is sufficient to justify inclusion of SOCs in accurate quantitative
simulations. On the other hand, the decay of fidelity due to NACs
is similar for both states. For the ground-state dynamics, NACs are
comparable in importance to SOCs, whereas for the excited state dynamics,
SOCs are clearly much more important.

\section{Conclusions\label{sec:Discussion-and-Conclusions}}

We have demonstrated that the fidelity criterion may detect disturbances
of the quantum dynamics due to SOCs, which may not be found if cruder
criteria, such as the extent of the population transfer between PESs,
are used. This observation is in accordance with our previous findings
on the nondiabaticity and nonadiabaticity of quantum dynamics.\cite{Zimmermann2010,Zimmermann2012}
It should be noted that fidelity is a nonspecific criterion and more
specific measures may be constructed when needed in order to attribute
the effect of couplings to either the electronic or nuclear dynamics.
This can be done, for example, by combination of the fidelity criterion
and surface population criterion.  However, the separation of the
nuclear and electronic effects is not always possible. Fidelity is
a rigorous single measure that can take into account both effects
simultaneously.

We found that the MSDR approximation of quantum fidelity remains accurate
when the perturbation is caused by SOCs. Based on our previous experience,
 the MSDR may fail, especially in cases where the underlying mixed
quantum\nobreakdash-classical dynamics does not approximate the quantum
dynamics reasonably well. This can happen, e.g., when tunneling is
important in the dynamics of the unperturbed Hamiltonian $\mathbf{\hat{H}}^{0}$.
In that case, any implementation of the MSDR, which is based on the
mixed quantum-classical dynamics, may not work. Other failures are
specific to current implementations of MSDR, where both the LMFD and
FSSH dynamics may be inaccurate due to the fact that all matrix elements
of the density matrix are attached to the same trajectory and evolve
on the same PES. A typical situation where approximations of this
kind fail occurs when regions of strong coupling are encountered repeatedly
and dynamics on PESs between these encounters differ substantially.
Surprisingly, in many cases the MSDR stays accurate despite the failure
of the underlying dynamics. This is due to the cancellation of errors
between the dynamics of $\mathbf{\hat{H}}^{0}$ and the (not explicitly
performed) dynamics of $\mathbf{\hat{H}}^{\epsilon}$.

Applications of the MSDR to the photoisomerization dynamics of CH\textsubscript{2}NH\textsubscript{2}\textsuperscript{+}
and to the collision of F + H\textsubscript{2} demonstrated that
our method may be used as a tool to decide which PESs and couplings
are important in the quantum dynamics, without the need to run quantum
dynamics itself. In contrast to the quantum dynamics, the MSDR does
not scale exponentially with the number of degrees of freedom, it
may be used on the fly and it does not require costly scans of PESs
or any other substantial prior knowledge of a system. (Note that there
exist several methods of quantum dynamics which may also be used on
the fly.\cite{Curchod2011,Wyatt2001,Lopreore2002} Nevertheless, they
are not currently widely used.) Finally, the MSDR is simple and its
current formulations may be very easily implemented into any code
for FSSH or Ehrenfest dynamics. 

\textit{Acknowledgement}

This research was supported by the Swiss NSF NCCR MUST (Molecular
Ultrafast Science \& Technology), Grant No. $200021\_124936/1$, and
by the EPFL. 

\bibliographystyle{apsrev}


\end{document}